\documentclass[11pt]{article}
\usepackage{amssymb,amsmath,amsfonts}
\usepackage{graphicx}
\usepackage{graphics}
\usepackage{eepic,epsfig}

\begin{document}

\title{OPTICAL TRANSITION RADIATION IN PRESENCE OF ACOUSTIC WAVES}
\author{A. R. Mkrtchyan, V. V. Parazian, A. A. Saharian \\
\textit{Institute of Applied Problems in Physics,}\\
\textit{25 Nersessian Street, 0014 Yerevan, Armenia}}
\maketitle

\begin{abstract}
Transition radiation from relativistic electrons is investigated in an
ultrasonic superlattice excited in a finite thickness plate. In the
quasi-classical approximation formulae are derived for the vector potential
of the electromagnetic field and for the spectral-angular distribution of
the radiation intensity. The acoustic waves generate new resonance peaks in
the spectral and angular distribution of the radiation intensity. The
heights of the peaks can be tuned by choosing the parameters of the acoustic
wave.
\end{abstract}

\bigskip


\bigskip


\section{Introduction}

Transition radiation is produced when a relativistic particle traverses an
inhomogeneous medium. Such radiation has a number of remarkable properties
and at present it has found many important applications. In particular,
optical transition radiation from metallic targets is widely used for the
measurement of transverse size, divergence and energy of electron and proton
beams (see, e.g. \cite{COTR} and references therein). The intensity of the
transition radiation can be increased considerably by using the interference
effects in periodic structures (for a review see Refs. \cite{TerMik}-\cite%
{Poty09}). From the point of view of controlling the parameters of the
high-energy electromagnetic processes in a medium, it is of interest to
investigate the influence of external fields, such as acoustic waves,
temperature gradient etc., on the corresponding characteristics. The
considerations of concrete processes, such as diffraction radiation \cite%
{MkrtDR}, parametric X-radiation \cite{Mkrt91}, channeling radiation \cite%
{Mkrt86}, electron-positron pair creation by high-energy photons \cite%
{Mkrt02}, bremsstrahlung of high-energy electrons \cite{Saha04Brem}, have
shown that the external fields can essentially change the angular-frequency
characteristics of the radiation intensities. Recently there has been broad
interest in compact crystalline undulators with periodically deformed
crystallographic planes as an efficient source of high energy photons \cite%
{Koro98} (for a review with a more complete list of references see \cite%
{Koro04}). In Ref. \cite{Grig98} we have considered the X-ray transition
radiation from ultrarelativistic electrons in an ultrasonic superlattice
excited in fused quartz plate. The radiation from a charged particle for a
semi-infinite laminated medium has been recently discussed in \cite{Grig09}.
The present paper deals with the optical transition radiation from
relativistic electrons in a plate in presence \ of ultrasonic vibrations.

The paper is organized as follows. In the next section we evaluate the
vector potential for the electromagnetic field in the quasi-classical
approximation. The radiation intensity is investigated in section \ref%
{sec:Radiation}. Special cases of the general formula are considered. In
section \ref{sec:Numeric} we present the results of the numerical evaluation
of the radiation intensity for the case of a plate made of fused quartz. The
main results of the paper are summarized in section \ref{sec:Conc}.

\section{Electromagnetic field}

\label{sec:Fields}

Consider the transition radiation emerging in the forward direction when a
charge $e$ (electron) moving with constant velocity $\mathbf{v}=v\mathbf{n}%
_{z}$ enters normally into a plate whose surfaces coincide with planes $z=-l$
and $z=0$. We assume that longitudinal ultrasonic vibrations are excited in
the plate along the normal to its surface (along the axis $z$), that form a
superlattice. The dielectric permittivity inside the plate we shall take in
the form
\begin{equation}
\varepsilon \left( z\right) =\varepsilon _{0}+\Delta \varepsilon \cos \left(
k_{s}z+\omega _{s}t+\varphi \right) ,  \label{epscond}
\end{equation}%
for $-l\leqslant z\leqslant 0$. In (\ref{epscond}), $\omega _{s}$, $k_{s}$
are the cyclic frequency and the wave number of the ultrasound, $\varphi $
is the initial phase. Under the condition $\nu _{s}l/v\ll 1$, with $\nu
_{s}=\omega _{s}/(2\pi )$, during the transit time of the electron the
dielectric constant in the superlattice is not notably changed. For
relativistic electrons and for the plate thickness $l\lesssim 1$ cm this
leads to the constraint $\nu _{s}\ll 10^{11}$ Hz. In the discussion below we
shall assume that the plate is immersed in a homogeneous medium with
dielectric permittivity $\varepsilon _{1}$.

We are interested in the radiation with frequencies $\omega $ satisfying the
condition $\omega \gg k_{s}c$. The presence of small parameter $%
k_{s}c/\omega $ allows one to use the quasi-classical approximation for the
evaluation of the radiation field. It is natural that in this case the plate
boundaries $z=-l$ and $z=0$ must be sufficiently smooth. This condition is
assumed to be observed \cite{TerMik} because the transition radiation on
these boundaries is formed in a zone with macroscopic length. For the
current density one has the expression
\begin{equation}
\mathbf{j}=ev\delta \left( x\right) \delta \left( y\right) \delta \left(
z-z\left( t\right) \right) \mathbf{n}_{z},\;z\left( t\right) =-l+v\left(
t-t_{0}\right) .  \label{currentmosh}
\end{equation}%
In the Lorentz gauge, the vector potential of the electromagnetic field can
be written as $\mathbf{A}=A\mathbf{n}_{z}$. This condition determines the
radiation polarization. The magnetic field is perpendicular to the plane
containing $\mathbf{n}_{z}$ and the photon wave vector.

In the quasi-classical approximation the Fourier component of the radiation
field,%
\begin{equation}
A( k_{\perp }, \omega , z) =\int_{-\infty }^{+\infty }A\left( \mathbf{r,}%
t\right) e^{i\left( \omega t-k_{1}x-k_{2}y\right) }\frac{dxdy}{\left( 2\pi
\right) ^{3}},\;\mathbf{k}_{\perp }=(k_{1},k_{2}),  \label{vectorpotdet}
\end{equation}%
in the region $z>0$ is determined by the expression \cite{TerMik}
\begin{equation}
A\left( k_{\perp },\omega ,z\right) =\frac{ieV}{4\pi ^{2}c}\sqrt{\frac{%
\varepsilon \left( z\right) }{k_{3}\left( z\right) }}\int_{-\infty
}^{+\infty }\exp \left[ i\omega t+i\int_{z\left( t\right)
}^{z}k_{3}(z^{\prime })dz^{\prime }\right] \frac{dt}{\sqrt{k_{3}(z\left(
t\right) )\varepsilon (z\left( t\right) )}}.  \label{Ak}
\end{equation}%
In formula (\ref{Ak}) we have
\begin{equation}
k_{3}(z)=\left\{
\begin{array}{ll}
k_{3}^{\left( 1\right) }, & z<-l,\;z>0, \\
k_{3}^{\left( 0\right) }+ak_{s}\cos \left( k_{s}z+\varphi _{1}\right) , &
-l<z<0,%
\end{array}%
\right.  \label{k3z}
\end{equation}%
and $\varphi _{1}=\omega _{s}t_{0}+\varphi $. We have assumed that $\Delta
\varepsilon $ is sufficiently small and the notations
\begin{equation}
k_{3}^{\left( j\right) }=\sqrt{\omega ^{2}\varepsilon _{j} /c^{2}-k_{\perp
}^{2}},\;j=0,1,\;a=\frac{\omega ^{2}\Delta \varepsilon }{2c^{2}k_{s}k_{3}^{%
\left( 0\right) }},  \label{defamek}
\end{equation}%
are introduced. Note that $a\sim \lambda _{s}\Delta \varepsilon /\lambda $,
where $\lambda _{s}$ is the wavelength of the acoustic wave and $\lambda $
is the wavelength of the radiated photon.

By using (\ref{k3z}) and the relation%
\begin{equation}
e^{ia\sin \tau }=\sum_{m=-\infty }^{+\infty }J_{m}\left( a\right) e^{im\tau
},  \label{besseldef}
\end{equation}%
where $J_{m}$ is the Bessel function of the first kind, after the
integration for the vector potential in the region $z>0$ one finds
\begin{eqnarray}
A( k_{\perp },\omega ,z ) &=&\frac{ie}{2\pi ^{2}c}\sqrt{\frac{\varepsilon
_{1}}{k_{3}^{(1)}}}\exp [i\omega (t_{0}+l/v)+ik_{3}^{\left( 1\right) }z]
\notag \\
&&\times \left\{ \frac{e^{ia\sin \left( \varphi _{1}\right) }}{\sqrt{%
k_{3}^{\left( 0\right) }\varepsilon _{0}}}\sum_{m=-\infty }^{+\infty
}J_{m}(a)e^{-il(\omega /v-k_{3m}^{\left( 0\right) })/2-im\varphi _{1}}\frac{%
\sin [l(\omega /v-k_{3m}^{\left( 0\right) })/2]}{\omega /v-k_{3m}^{\left(
0\right) }}\right.  \notag \\
&&+\left. \frac{e^{i\phi _{1}/2}\sin (\phi _{1}/2)-e^{i\phi _{1}-il(\omega
/v-k_{3}^{(1)})/2}\sin [l(\omega /v-k_{3}^{(1)})/2]}{\sqrt{%
k_{3}^{(1)}\varepsilon _{1}}(\omega /v-k_{3}^{(1)})}\right\} .  \label{Ak1}
\end{eqnarray}%
Here we have used the notations%
\begin{eqnarray}
k_{3m}^{\left( 0\right) } &=&k_{3}^{\left( 0\right) }+mk_{s},  \notag \\
\phi _{1} &=&(k_{3}^{\left( 0\right) }-k_{3}^{(1)})l+a\left[ \sin \left(
\varphi _{1}\right) +\sin \left( k_{s}l-\varphi _{1}\right) \right] .
\label{k3m0}
\end{eqnarray}%
The corresponding expressions for the electric and magnetic fields are found
by using the standard formulae of electrodynamics.

\section{Radiation intensity}

\label{sec:Radiation}

Expression (\ref{Ak1}) may be used in calculations of the radiation
intensity in the region $z>0$. We shall denote by $\theta $ the angle
between the momentum of the radiated photon and the axis $z$. The energy
radiated during the electron transit time in the range of frequencies $%
d\omega $ and angles $d\theta $ is determined by the relation \cite{TerMik}%
\begin{equation}
I\left( \omega ,\theta \right) d\omega d\theta =\left( 2\pi \right) ^{3}%
\frac{\varepsilon _{1}^{3/2}\omega ^{4}}{c^{3}}\sin ^{3}\theta \cos
^{2}\theta \left\vert A(k_{\perp },\omega ,z\gg l)\right\vert ^{2}d\omega
d\theta ,  \label{specangdistermik}
\end{equation}%
where $k_{\perp }=(\omega \sqrt{\varepsilon _{1}}/c)\sin \theta $. \ For the
beam of relativistic electrons the expression $I\left( \omega ,\theta
\right) $ should be averaged over the phase $\varphi _{1}$ of particle
flight into the plate. After this procedure, for the spectral-angular
density of the radiated energy in the angular region $\sin \theta <\sqrt{%
\varepsilon _{0}/\varepsilon _{1}}$ we find%
\begin{eqnarray}
I\left( \omega ,\theta \right)  &=&\frac{2e^{2}\beta _{1}^{2}}{\pi c\sqrt{%
\varepsilon _{1}}}\sin ^{3}\theta \sum_{m=-\infty }^{+\infty
}J_{m}^{2}\left( \frac{\omega \Delta \varepsilon }{2ck_{s}\sqrt{\varepsilon
_{0}-\varepsilon _{1}\sin ^{2}\theta }}\right)   \notag \\
&&\times \sin ^{2}[\frac{l\omega }{2v}(1-\beta _{1}\sqrt{\varepsilon
_{0}/\varepsilon _{1}-\sin ^{2}\theta }+mk_{s}v/\omega )]  \notag \\
&&\times \left[ \frac{\sqrt{\varepsilon _{1}/\varepsilon _{0}}(\cos \theta /%
\sqrt{\varepsilon _{0}/\varepsilon _{1}-\sin ^{2}\theta })^{1/2}}{1-\beta
_{1}\sqrt{\varepsilon _{0}/\varepsilon _{1}-\sin ^{2}\theta }+mk_{s}v/\omega
}-\frac{1}{1-\beta _{1}\cos \theta }\right] ^{2},  \label{Iom1}
\end{eqnarray}%
where $\beta _{1}=v\sqrt{\varepsilon _{1}}/c$ and $\varepsilon _{0}$, $%
\varepsilon _{1}$ are functions of $\omega $. Note that for relativistic
electrons in this formula we have $k_{s}v/\omega \approx \lambda /\lambda
_{s}$. For a given value of the radiation wavelength $\lambda =2\pi c/\omega
$, the peaks $\theta =\theta _{m}(\lambda )$ in the angular distribution of
the radiation intensity are determined from the equation%
\begin{equation}
\sin ^{2}\theta _{m}(\lambda )=\frac{\varepsilon _{0}}{\varepsilon _{1}}-%
\frac{1}{\varepsilon _{1}}\left( \frac{c}{v}+m\frac{\lambda }{\lambda _{s}}%
\right) ^{2}.  \label{angPeaks}
\end{equation}%
The width of these peaks is of the order $(c/v+m\lambda /\lambda
_{s})\lambda /[l\sin (2\theta _{m})]$. For the radiation inside
the plate the angular peaks in the radiation intensity are located
at the outgoing photon angles $\theta =\theta _{m}^{(0)}(\lambda
)=\arccos [(c/v+m\lambda /\lambda _{s})/\sqrt{\varepsilon _{0}}]$.
After the refraction on the boundary at $z=0$, the corresponding
outgoing angle for the radiation in the region $z>0$ is determined
by the relation $\sin \theta =\sqrt{\varepsilon _{0}/\varepsilon
_{1}}\sin [\theta _{m}^{(0)}(\lambda )]$ which coincides with
$\theta _{m}(\lambda )$ defined by (\ref{angPeaks}).

The location of the peaks does not depend on the amplitude of the acoustic
wave. The amplitude determines the heights of the peaks. The corresponding
dependence is described by the Bessel functions in (\ref{Iom1}). Under the
assumption $\lambda /\lambda _{s}\ll 1$, made above, at the peaks of the
radiation corresponding to (\ref{angPeaks}) the argument of the Bessel
functions is approximately equal $\Delta \varepsilon \lambda _{s}/(2\lambda
) $. Hence, for given values of $\lambda $, $\lambda _{s}$, and $m$, the
dependence of the peaks heights on the acoustic wave amplitude is described
by the function $J_{m}^{2}(\Delta \varepsilon \lambda _{s}/(2\lambda ))$. In
particular, the maximum height of the peak is obtained for the amplitude of
the acoustic wave corresponding to $\Delta \varepsilon \approx 2x_{m}\lambda
/\lambda _{s}$, where $x=x_{m}$ corresponds to the maximum of the function $%
J_{m}^{2}(x)$. For $m=0,1,2$ one has $x_{m}=0,1.84,3.05$, respectively. From
the well-known properties of the Bessel functions it follows that the height
of the central peak corresponding to $m=0$ decreases with increasing $\Delta
\varepsilon $ in the range $0\leqslant \Delta \varepsilon \leqslant
4.76\lambda /\lambda _{s}$. The heights of the peaks generated by the
acoustic wave increase with increasing $\Delta \varepsilon $ in the range $%
0\leqslant \Delta \varepsilon \leqslant 3.68\lambda /\lambda _{s}$ and $%
0\leqslant \Delta \varepsilon \leqslant 6.1\lambda /\lambda _{s}$ for $m=1$
and $m=2$ respectively. These features of the dependence of the radiation
intensity on the acoustic wave amplitude will be illustrated below by a
numerical example.

Let us consider special cases of formula (\ref{Iom1}). For X-ray transition
radiation one has $\varepsilon _{0}=1-\omega _{\mathrm{pl}}^{2}/\omega ^{2}$
and formula (\ref{Iom1}) with $\varepsilon _{1}=1$ is reduced to the one
derived in Ref. \cite{Grig98}. In the absence of the acoustic wave we have $%
\Delta \varepsilon =0$ and (\ref{Iom1}) takes the form
\begin{eqnarray}
I_{0}\left( \omega ,\theta \right)  &=&\frac{2e^{2}\beta _{1}^{2}}{\pi c%
\sqrt{\varepsilon _{1}}}\sin ^{3}\theta \sin ^{2}[\frac{l\omega }{2v}%
(1-\beta _{1}\sqrt{\varepsilon _{0}/\varepsilon _{1}-\sin ^{2}\theta })]
\notag \\
&&\times \left[ \frac{\sqrt{\varepsilon _{1}/\varepsilon _{0}}(\cos \theta /%
\sqrt{\varepsilon _{0}/\varepsilon _{1}-\sin ^{2}\theta })^{1/2}}{1-\beta
_{1}\sqrt{\varepsilon _{0}/\varepsilon _{1}-\sin ^{2}\theta }}-\frac{1}{%
1-\beta _{1}\cos \theta }\right] ^{2}.  \label{Iom0}
\end{eqnarray}%
For a given $\omega $ the location of the peaks in the angular distribution
of the radiation intensity is determined from the relation $\sin ^{2}\theta
=\varepsilon _{0}/\varepsilon _{1}-\beta _{1}^{-2}$. It is of interest to
note that the ongoing photon angle inside the plate corresponding to this $%
\theta $ is equal to the Cherenkov angle for the material of the plate: $%
\sin \theta =\sqrt{\varepsilon _{1}/\varepsilon _{0}}\sin \theta _{\mathrm{C}%
}$, where $\cos \theta _{\mathrm{C}}=c/(v\sqrt{\varepsilon _{0}})$. As we
see, for the appearance of the peaks in the forward direction we need to
have the condition $c^{2}/v^{2}\leqslant \varepsilon _{0}\leqslant
\varepsilon _{1}+c^{2}/v^{2}$. Otherwise, Cherenkov radiation emitted inside
the plate will be reflected from the boundary. Formula (\ref{Iom0}) gives
the quasi-classical approximation for the radiation intensity emitted by a
charge flying into a dielectric plate in the direction of normal to its
surface. The exact expression for the radiation intensity in this problem is
given by Pafomov formula \cite{Pafo69} (see also Refs. \cite{TerMik}-\cite%
{Poty09}). A detailed consideration of optical transition radiation
described by Pafomov formula is recently given in Ref. \cite{Hrmo00}. In
this paper, in discussing the transition radiation in a dielectric plate, as
a transparent material is taken glass. Numerical results based on Pafomov
formula are given for the refractive index $n=1.5$. For this value of the
refractive index Cherenkov radiation emitted in the plate is completely
reflected from the boundary (see the condition given above). For a
transparent material in the over-threshold case and under the condition $%
l\omega /(2v)\gg 1$ the dominant contribution comes from the term in Pafomov
formula with the resonant factor $\sin ^{2}[l\omega x/(2v)]/x^{2}$, where $%
x=1-\beta _{1}\sqrt{\varepsilon _{0}/\varepsilon _{1}-\sin ^{2}\theta }$.
Now, for simplicity considering the case $\varepsilon _{1}=1$, it can be
seen that for a relativistic particle with $1-v/c\ll 1$ the radiation
intensity near the Cherenkov peaks, corresponding to $\sin \theta =\sqrt{%
\varepsilon _{0}-c^{2}/v^{2}}$, is well approximated by formula (\ref{Iom0})
(we have checked this both analytically and numerically (see below) for
fused quartz with the dispersion law given in the next section).

The Cherenkov peaks coincide with the $m=0$ peaks in the presence of the
acoustic wave. As it follows from (\ref{Iom1}), the height of the peaks is
reduced by the factor $J_{0}^{2}(a)$. For the parameters of the acoustic
wave taken in accordance with $a=j_{0,s}$, $s=1,2,\ldots $, where $z=$ $%
j_{0,s}$ are the zeros of the function $J_{0}(z)$, the corresponding
radiation intensity vanishes. In the absence of acoustic waves the frequency
range in which the peaks appear is determined by the condition $\varepsilon
_{0}\leqslant \varepsilon _{1}+(c/v)^{2}$. When the acoustic waves are
present the corresponding condition takes the form $\varepsilon
_{0}\leqslant \varepsilon _{1}+(c/v+m\lambda /\lambda _{s})^{2}$. The latter
condition is less restrictive. In particular, we can have a situation when
in the absence of the acoustic waves there are no peaks in a given frequency
range and they appear in the presence of acoustic waves. Examples for this
type of situation will be given below.

In the limit $l\rightarrow \infty $, by using the formula $%
\lim_{x\rightarrow \infty }\sin ^{2}\alpha x/x=\pi \alpha ^{2}\delta (\alpha
)$, for the spectral-angular density of the radiation energy per unit length
we find%
\begin{eqnarray}
I_{\infty }\left( \omega ,\theta \right) &=&\lim_{l\rightarrow \infty
}I\left( \omega ,\theta \right) /l=\frac{e^{2}\varepsilon _{1}^{3/2}\omega
^{2}}{2c^{3}\varepsilon _{0}}\sum_{m=-\infty }^{+\infty }J_{m}^{2}\left(
\frac{\omega \Delta \varepsilon }{2ck_{s}\sqrt{\varepsilon _{0}-\varepsilon
_{1}\sin ^{2}\theta }}\right)  \notag \\
&&\times \frac{\sin ^{3}\theta \cos \theta }{\sqrt{\varepsilon
_{0}/\varepsilon _{1}-\sin ^{2}\theta }}\delta \left( \frac{\omega }{2v}%
(1-\beta _{1}\sqrt{\varepsilon _{0}/\varepsilon _{1}-\sin ^{2}\theta })+%
\frac{1}{2}mk_{s}\right) .  \label{Iominf}
\end{eqnarray}%
As we see, in this case the radiation for a given $\theta $ has discrete
spectrum determined by the relation

\begin{equation}
\omega =\frac{-mk_{s}v}{1-\beta _{1}\sqrt{\varepsilon _{0}/\varepsilon
_{1}-\sin ^{2}\theta }}.  \label{omn}
\end{equation}%
This relation directly follows from the periodicity properties of the
problem. The spectral distribution of the radiation intensity is obtained
from (\ref{Iominf}) by the integration over $\theta $ and in the case $%
\varepsilon _{1}=\varepsilon _{0}$ coincides with the result given in \cite%
{TerMik}.

In the discussion above we have assumed that the particle moves with
constant velocity along the direction perpendicular to the plate. In reality
the particle trajectory will be affected by the interaction with atoms of
medium (multiple scattering). It is known (see, for example, \cite{TerMik})
that the multiple scattering does not suppress the X-ray transition
radiation. Moreover, it leads to the widening of the frequency range. An
exact treatment for the influence of the multiple scattering on the
intensity of radiation in the problem under consideration requires a
separate consideration. We expect that the multiple scattering will result
in the broadening of the radiation angular distribution for the peaks
described above. By taking into account that the change in the medium
density induced by acoustic wave is small, in the first approximation we can
estimate this broadening by using the corresponding results for Cherenkov
radiation in a homogeneous medium given in Ref. \cite{Gric06}. In this way,
for the ratio of the angular broadening due to the multiple scattering to
the angular width of the peaks estimated above we find $\approx l\sigma
_{1}/(\pi \varepsilon _{0})$, where $\sigma _{1}$ is defined by the mean
squared angle of multiple scattering per unit length of the particle
trajectory. Note that the characteristics of the peaks discussed above are
determined by the relative velocity $v/c$ and for $v/c\approx 1$ they are
relatively unsensitive to the energy of the particle, whereas the parameter $%
\sigma _{1}$ is decreasing with increasing energy. Hence, by increasing the
particle energy we can reduce the relative role of the effects related to
the multiple scattering.

\section{Numerical evaluation}

\label{sec:Numeric}

Here we are interested in the optical transition radiation. For the
numerical evaluation of the radiation intensity the material of the plate
should be specified. We assume that the plate is made of fused quartz. For
the dielectric permittivity of fused quartz we use the Sellmeier dispersion
formula%
\begin{equation}
\varepsilon _{0}=1+\sum_{i=1}^{3}\frac{a_{i}\lambda ^{2}}{\lambda
^{2}-l_{i}^{2}}  \label{sellmeier}
\end{equation}%
where $a_{1}=0.6961663$, $a_{2}=0.4079426$, $a_{3}=0.8974794$, $%
l_{1}=0.0684043$, $l_{2}=0.1162414$, $l_{3}=9.896161$, and $\lambda $ is the
wavelength of the radiation measured in micrometers. Formula (\ref{sellmeier}%
) well describes the dispersion properties of fused quartz in the range $0.2%
\mathrm{\mu m}\leqslant \lambda \leqslant 6.7\mathrm{\mu m}$. Note that in
this spectral region fused quartz is very weakly absorbing. For fused quartz
the velocity of longitudinal ultrasonic vibrations is $\omega
_{s}/k_{s}\approx 5.6\times 10^{5}\;$cm/s. In figures below we have plotted
the spectral-angular density of the radiation intensity, $I(\omega ,\theta
)/\hbar $, for the energy of electrons equal to $2\;$MeV and for the plate
thickness $l=0.5$ cm. For the oscillation amplitude we have taken the value $%
\Delta n/n_{0}=0.05$, where $n_{0}$ is the number of electrons per unit
volume for fused quartz.

\begin{figure}[tbph]
\begin{center}
\begin{tabular}{cc}
\epsfig{figure=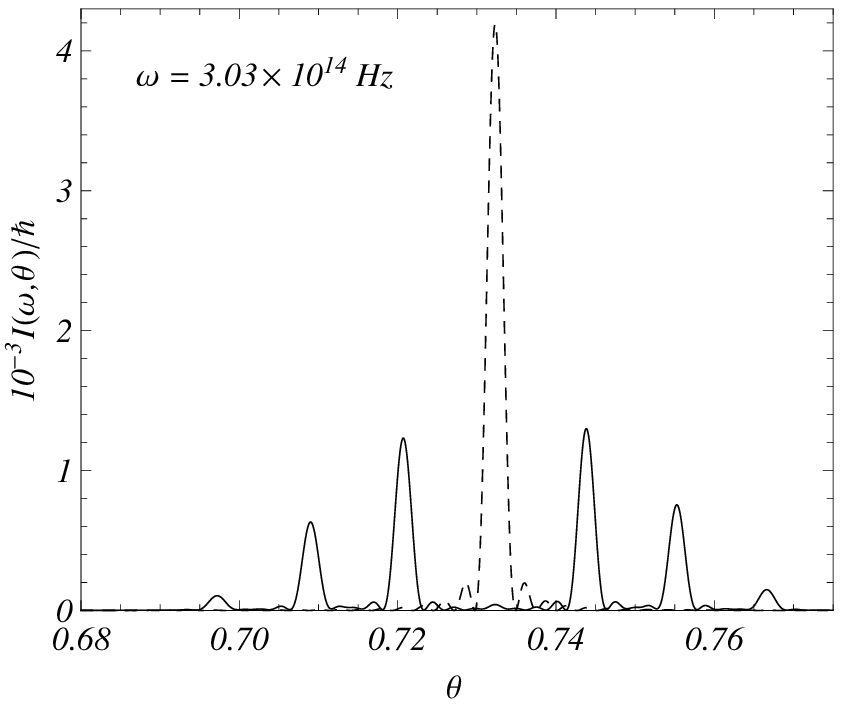,width=7.cm,height=6.cm} & \quad %
\epsfig{figure=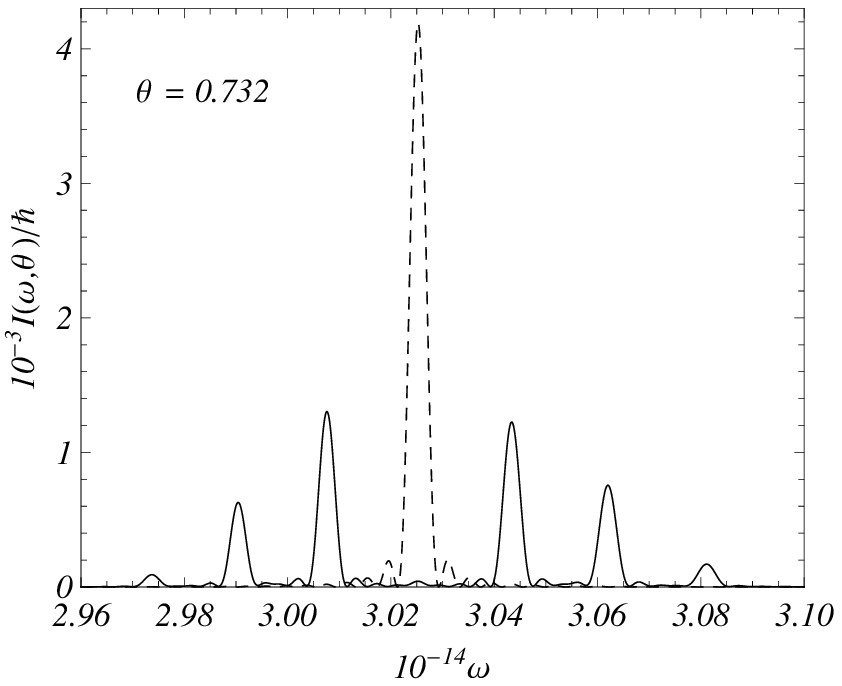,width=7.cm,height=6.cm}%
\end{tabular}%
\end{center}
\caption{The angular (left panel) and spectral (right panel) distributions
of the radiation intensity. The dashed curves correspond to the transition
radiation when the acoustic wave is absent. For the left panel $\protect%
\omega =3.03\times 10^{14}$ Hz and for the right panel $\protect\theta %
=0.732 $. The frequency of the acoustic wave is equal to 5 MHz. }
\label{fig1}
\end{figure}

As it is seen from the graphs given below the presence of the acoustic wave
leads to the appearance of new peaks in both angular and spectral
distributions of the radiation intensity. The height of the peaks can be
tuned by choosing the parameters of the acoustic wave. In particular, the
peak in the radiation intensity which is present in the absence of the
acoustic wave is reduced by the factor $J_{0}^{2}(a)$, where $a$ is the
argument of the Bessel function in (\ref{Iom1}). This peak can be completely
removed by taking the parameters of the acoustic wave in such a way to have $%
a=j_{0,s}$, $s=1,2,3,\ldots $, where $z=j_{0,s}$ are the zeroes of the
function $J_{0}(z)$. In a similar way other peaks can be suppressed by
tuning the acoustic wave parameters. The graphs in figure \ref{fig1} (full
curves) are plotted for the frequency of acoustic wave $\nu _{s}=5$ MHz. The
dashed curves correspond to the radiation intensity in the absence of the
acoustic wave. The left/right panel presents the angular/spectral
distribution of the radiation intensity for a given frequency/angle. The
parameters are chosen in a way to suppress the peak in the absence of the
acoustic wave.

\begin{figure}[tbph]
\begin{center}
\begin{tabular}{cc}
\epsfig{figure=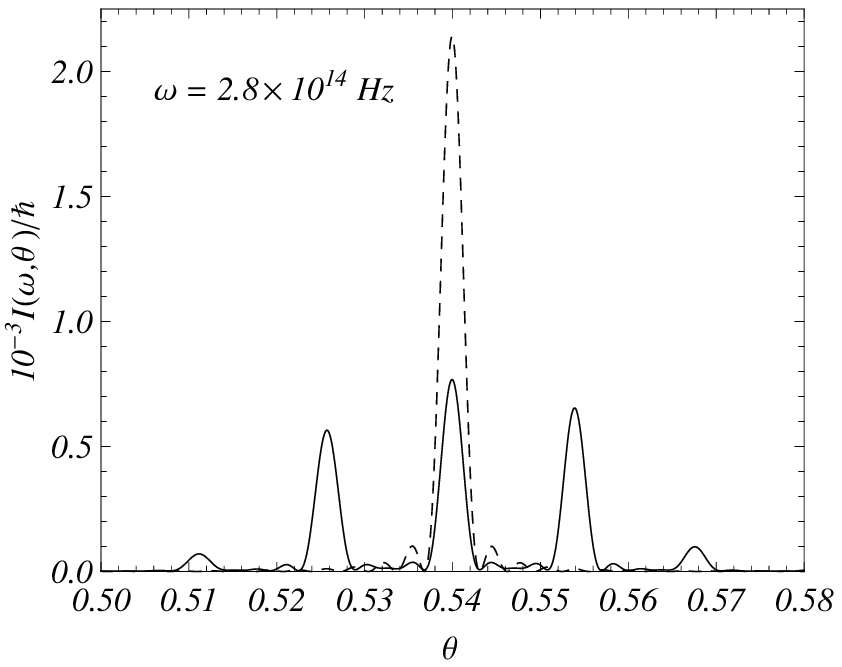,width=7.cm,height=6.cm} & \quad %
\epsfig{figure=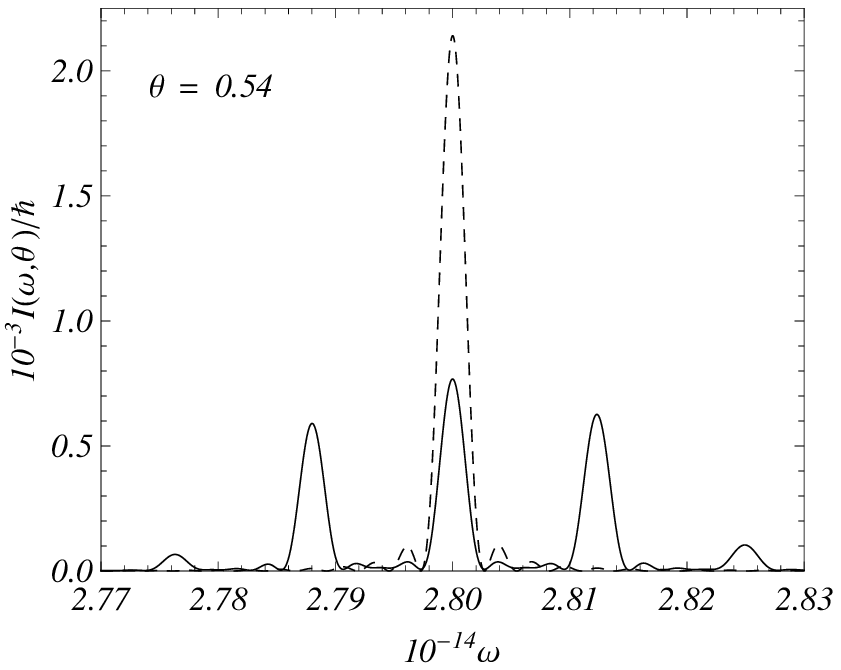,width=7.cm,height=6.cm}%
\end{tabular}%
\end{center}
\caption{The same as in figure \protect\ref{fig1} for $\protect\omega %
=2.8\times 10^{14}$ Hz (left panel) and for $\protect\theta =0.54$ (right
panel). }
\label{fig2}
\end{figure}

\begin{figure}[tbph]
\begin{center}
\epsfig{figure=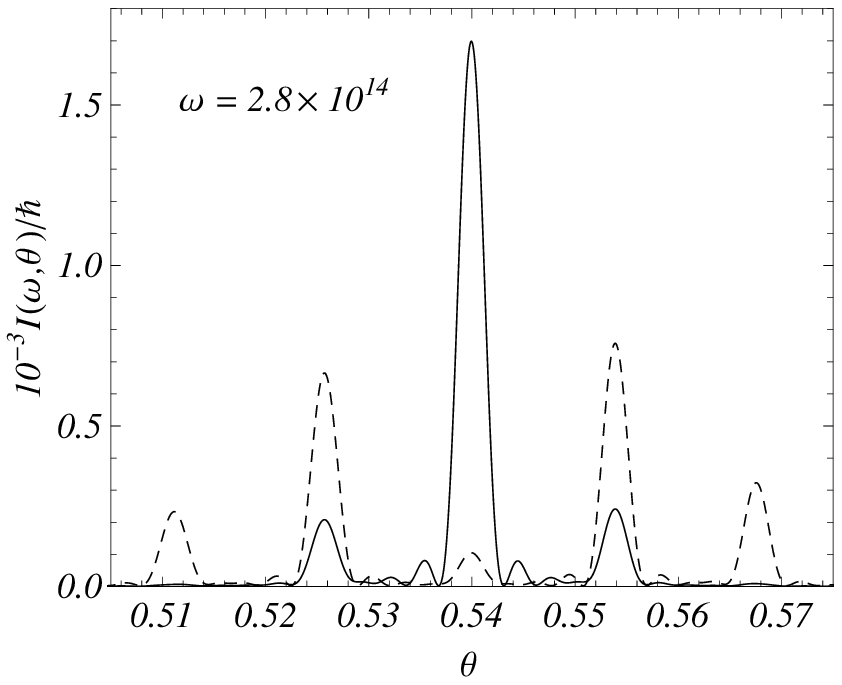,width=7.cm,height=6.cm}
\end{center}
\caption{The same as in figure \protect\ref{fig2} for the amplitudes of
acoustic wave corresponding to $\Delta n/n_{0}=0.025$ (full curve) and $%
\Delta n/n_{0}=0.075$ (dashed curve).}
\label{fig3}
\end{figure}

An example when the central peak is reduced partially is presented in figure %
\ref{fig2}. In order to illustrate the dependence of the radiation intensity
on the amplitude of the ultrasonic vibrations, for this case, in figure \ref%
{fig3} we plotted the angular distribution of the radiation intensity for $%
\Delta n/n_{0}=0.025$ (full curve) and $\Delta n/n_{0}=0.075$ (dashed
curve). As it is seen, the height of the central peak decreases with
increasing amplitude, whereas the heights for the peaks generated by the
acoustic wave increase. This is in agreement with the features described in
the previous section. For the values of the parameters corresponding to
figures \ref{fig2} and \ref{fig3} we have also evaluated the radiation
intensity near the Cherenkov peak in the absence of acoustic wave by using
Pafomov formula. The corresponding results coincide with those evaluated
within the framework of quasi-classical approximation with the relative
accuracy more than 99\%.

\begin{figure}[tbph]
\begin{center}
\begin{tabular}{cc}
\epsfig{figure=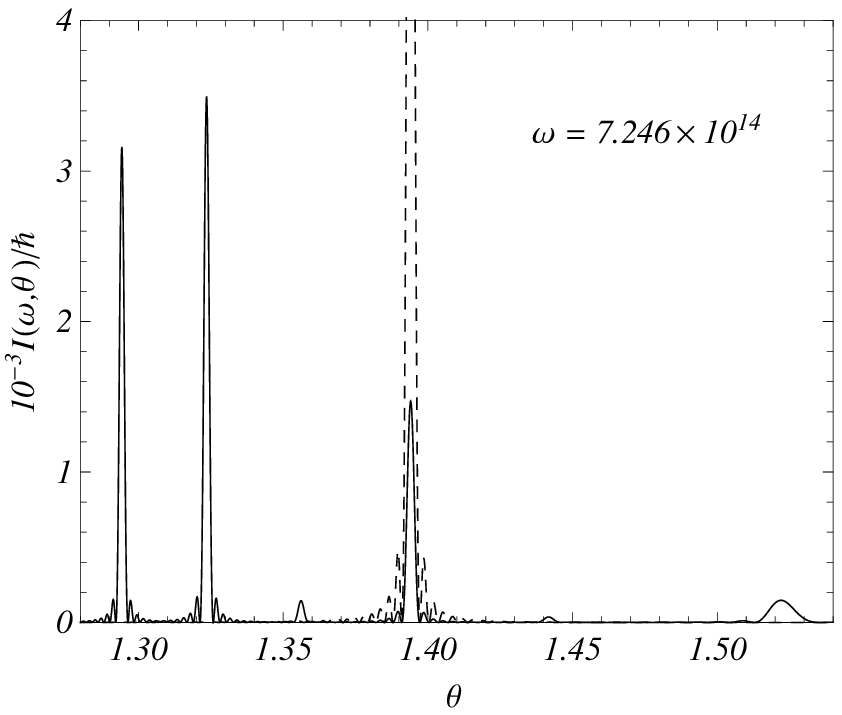,width=7.cm,height=6.cm} & \quad %
\epsfig{figure=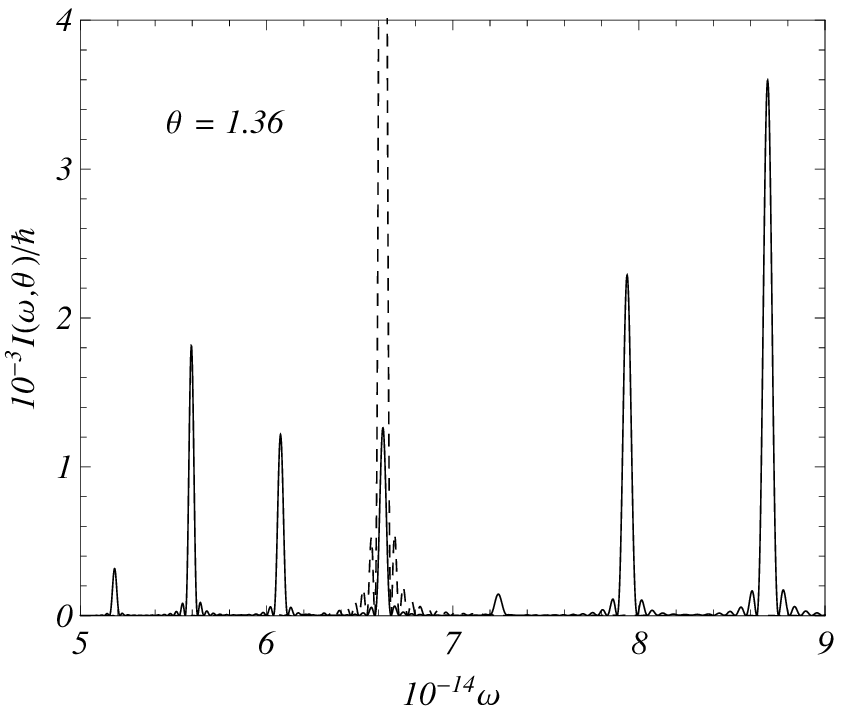,width=7.cm,height=6.cm}%
\end{tabular}%
\end{center}
\caption{The same as in figure \protect\ref{fig1} for $\protect\omega %
=7.246\times 10^{14}$ Hz (left panel) and for $\protect\theta =1.36$ (right
panel). The frequency of the acoustic wave is equal to 15 MHz. }
\label{fig4}
\end{figure}

\begin{figure}[tbph]
\begin{center}
\epsfig{figure=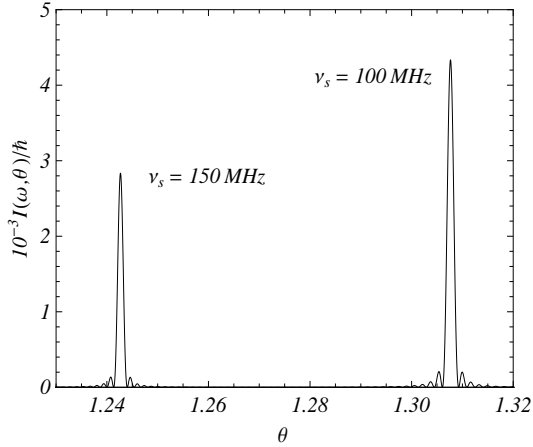,width=7.cm,height=6.cm}
\end{center}
\caption{The angular distribution of the radiation intensity for $\protect%
\omega =1\times 10^{15}$ Hz. The frequency of the acoustic wave is equal to
100 MHz and 150 MHz. }
\label{fig5}
\end{figure}

In figure \ref{fig4} we illustrate the suppression of the $m=\pm 1$ peaks in
the case of acoustic wave with frequency $\nu _{s}=15$ MHz. For the left
(right) panel the height of the peak in the absence of the acoustic wave
(dashed curves) is $\approx 9.5$ ($\approx 11.4$). In the absence of
acoustic wave and for the energy of electron $\geqslant 2$ MeV the peaks in
the forward direction appear in the range $\omega <9.62\times 10^{14}$. In
figure \ref{fig5} we give an example when there is no peak in the absence of
acoustic wave and the peaks appear when the acoustic wave is excited. The
presented graphs correspond to the frequencies of acoustic wave 100 MHz and
150 MHz and to the frequency of the radiation $\omega =1\times 10^{15}$ Hz.
Note that for all presented examples we have $\lambda \ll \lambda _{s}$ and
the condition for the validity of the quasi-classical approximation is
satisfied.

\section{Conclusion}

\label{sec:Conc}

In the present paper we have investigated the transition radiation from an
electron in a plate in the presence of acoustic waves. In the
quasi-classical approximation we have derived formulae for the
electromagnetic field and the radiation intensity in the forward direction.
The spectral-angular density of the radiated energy is given by formula (\ref%
{Iom1}). The numerical examples are given for a plate of fused quartz. These
results show that the acoustic waves allow to control the parameters of the
radiation. In particular, new resonance peaks appear in the spectral-angular
distribution of the radiation intensity. The height of the peaks can be
tuned by choosing the parameters of the acoustic wave.

\end{document}